\documentstyle{l-aa}
\begin{document}
\thesaurus{11 (11.11.1; 11.16.1; 11.09.1 NGC 6181)}
\title{A Ring-like Zone of Strong Radial Gas Motions in the Disk of NGC 6181
\thanks{Based on observations collected with the 6m telescope
at the Special Astrophysical Observatory (SAO) of the
Russian Academy of Sciences (RAS).}}
\author{O. K. Sil'chenko \inst{1} \and A. V. Zasov \inst{1}
\and A. N. Burenkov \inst{2} \and J. Boulesteix \inst{3}}
\offprints{O. K. Sil'chenko}
\institute{
Sternberg Astronomical Institute, University av. 13,
Moscow 119899, Russia
\and
Special Astrophysical Observatory, Nizhnij Arkhyz, Karachaj-Cherkess
Republic, 357147 Russia
\and Observatoire de Marseille, 2 Place Le Verrier,
F-13248 Marseille Cedex 04, France}
\maketitle
% \markboth
% \maintitlerunninghead{Radial Gas Motions in NGC 6181}
% \authorrunninghead{O. K. Sil'chenko et al.}
\begin{abstract}

The Sc galaxy NGC 6181 was observed at the 6m telescope of SAO RAS
with the scanning Perot-Fabry interferometer in the H$_\alpha$
emission line and at the 1m telescope of SAO RAS in $BVRI$ broadband filters
with CCD. Subtraction of the mean circular rotation curve from the
two-dimensional velocity field has revealed a ring-like zone with a
diameter about of 2 kpc where strong radial gas motions are present.
The form of the ring is almost perfectly circular in the plane of the
galaxy. It is located closer to the center than the beginning of the
well-defined spiral structure, but outside of the central bulge-dominated
region. The detected radial velocity reduced to the plane of the
galaxy is about 100 $km\cdot s^{-1}$ and probably is azimuthally
dependent. The very inner region of the galaxy, $r<3\arcsec$ or 0.5
kpc, shows a turn of the dynamical major axis by about $30^o$. Central
continuum isophotes are also twisted which suggests the
presence of small nuclear bar .

\keywords{Galaxies: spirals, kinematics -
Scanning Perot-Fabry interferometry}

\end{abstract}

\section{Introduction}
NGC 6181 is an isolated late-type giant spiral galaxy
whose high surface brightness
of gas emission and inclination of about $60^o$ are favorable
for a detailed kinematical study. Global parameters of the galaxy are
listed in Table 1 being taken basically from LEDA (Lyon-Meudon
Extragalactic Database). The first kinematical investigation of NGC
6181 was undertaken nearly 30 years ago by Burbidge et al. (1965).
They obtained three spectral long-slit
cross-sections of the galaxy and found line-of-sight velocity
distributions to be quite asymmetrical. They  concluded that NGC 6181
is not an axisymmetric galaxy being intermediate between a barred and
a normal spiral. The form of thin dust lanes in the center of NGC 6181
gave some support to this hypothesis. In addition, they noted that
"the structural center of the galaxy is not the center of the velocity
distribution", because the systemic velocities determined from the
outer and the inner parts of line-of-sight velocity curves disagreed.

Our group (Afanasiev et al. 1992) has repeated a long-slit
kinematical study of NGC 6181 at the 6m telescope of the Special Astrophysical
Observatory (SAO RAN) and fully confirmed the unusual asymmetrical
character of the one-dimensional line-of-sight velocity distributions.
It has been proposed that it is the southern part of the galaxy which
reveals strong non-circular gas motions up to distances of about
20\arcsec\ from the center. But it became evident that the obtaining of
the two-dimensional velocity field is necessary to clarify the situation.

\begin{table}
\caption[ ] {Global parameters of NGC 6181}
% %\begin{center}
\begin{flushleft}
\begin{tabular}{ll}
\hline\noalign{\smallskip}
% % &  \\
Hubble type & SAB(rs)c  \\
$R_{25}$ & 12.3 kpc \\
$B_T^0$ & 11.7  \\
$M_B$ & --20.93  \\
$V_r(radio)$ & 2374 $km\cdot s^{-1}$ \\
$V_{gal.std-of-rest}$ & 2492 $km\cdot s^{-1}$  \\
Distance & 33.6 Mpc ($H_0$=75 $km\cdot s^{-1}\cdot Mpc^{-1}$) \\
Inclination & $66.6^o$  \\
\it{MA}$_{phot}$ & $175^o$  \\
\hline
\end{tabular}
\end{flushleft}
\end{table}

\section{Observations}

The two-dimensional velocity field of NGC 6181 was obtained at the 6m
telescope on September 24, 1993. The scanning Perot-Fabry
interferometer was installed inside of the pupil plane of a focal
reducer which was attached to the F/4 prime focus of the telescope.
An intensified photon counting system (IPCS) $512\times 512$ was used as
the detector. Instrument parameters are given in Table 2.

\begin{table}
\caption[ ] {Scanning Perot-Fabry observations parameters}
\begin{flushleft}
\begin{tabular}{ll}
\hline\noalign{\smallskip}
Input telescope beam & F/4 \\
Output camera beam & F/2.4 \\
Number of pixels & $512\times 512$ \\
Pixel scale & 0.35\arcsec \\
Field & $3\arcmin \times 3\arcmin$ \\
Filter wavelength & $\lambda_f$=6620 \AA \\
Filter FWHM & 10 \AA \\
Galaxy H$_\alpha$ wavelength & $\lambda_g$=6615 \AA \\
Filter transmission at $\lambda_g$ & 35 \% \\
Etalon interference order & 501 at 6562.8\AA \\
Free spectral range at $\lambda_g$ & 603 $km\cdot s^{-1}$ \\
Finesse at $\lambda_g$ & 18 \\
Number of scanned channels & 32 \\
Channel step & 18.84 $km\cdot s^{-1}$ \\
Number of scanned cycles & 7 \\
Basic exposure time & 20 s \\
Total exposure time & 4480 s \\
Calibration line & Neon 6598.98 \AA \\
Seeing & ${< 1.5\arcsec}$ \\
\hline
\end{tabular}
\end{flushleft}
\end{table}

The "Image-interferometry method" was described in detail earlier
(Boulesteix et al. 1983, Amram et al. 1991). Observational data
(galaxy and wavelength calibration data) were converted into cubes of
32 images ($256\times256$), with the linear scale being 0.70 arcsec/px
and a spectral resolution of about 2--2.5 channels (40--50 $km\cdot s^{-1}$).
A reduction of observational data (correction for phase shifting,
subtraction of night-sky emission spectrum, construction of velocity
map, a.s.o.) was done by using standard methods, and the Perot-Fabry
reducing software ADHOC developed at Marseille Observatory
(Boulesteix 1993) was used.

The continuum subtracted H$_\alpha$ image of NGC 6181 (Fig. 1) reveals that
the H$_\alpha$ emission is rather strong over the whole galactic disk.
So the high quality of the two-dimensional velocity distribution obtained
allows to get more detailed information about gas motions in the disk
of this galaxy. A sub-cube ($210 \,px\times 185 \,px \times 32 \,channels$),
containing the main fraction of H$_\alpha$ emission from the galaxy,
was extracted from a reduced data cube. This smaller cube was used for
the analysis of the velocity field.

\begin{figure}
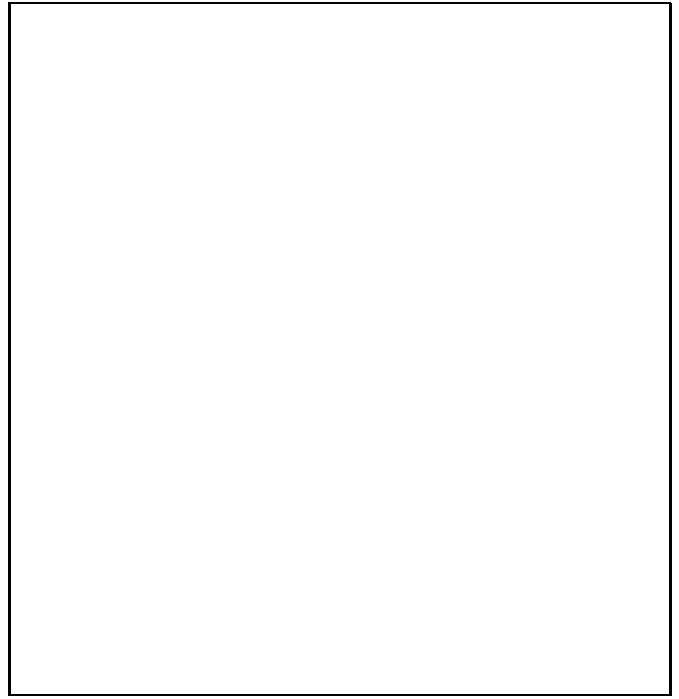

% Fig 1
\picplace{9.2 cm}
\caption[ ]{Monochromatic H$_\alpha$ emission map of NCC 6181. The cross
marks the position of the nucleus.}
\end{figure}

\begin{table}
\caption[ ] {Direct image observations}
% \begin{center}
\begin{flushleft}
\begin{tabular}{cccc}
\hline\noalign{\smallskip}
Date & Filter & Exposure time & Zenithal distance \\
\hline\noalign{\smallskip}
7/8.06.94   & $I$ & 600 s & $49^o$ \\
7/8.06.94   & $I$ & 100 s & $52^o$ \\
8/9.06.94   & $I$ & 600 s & $50^o$ \\
8/9.06.94   & $R$ & 200 s & $52^o$ \\
8/9.06.94   & $R$ & 200 s & $53^o$ \\
8/9.06.94   & $R$ & 200 s & $54^o$ \\
10/11.07.94 & $B$ & 300 s & $57^o$ \\
10/11.07.94 & $V$ & 300 s & $59^o$ \\
10/11.07.94 & $B$ & 300 s & $60^o$ \\
10/11.07.94 & $V$ & 300 s & $61^o$ \\
10/11.07.94 & $I$ & 300 s & $63^o$ \\
\hline
\end{tabular}
% \end{center}
\end{flushleft}
\end{table}

Direct images of NGC 6181 were obtained at the telescope Zeiss--1000
of SAO RAN. Eleven frames of the galaxy have been derived with a CCD
camera through the $B$, $V$, $R$ and $I$ filters of Johnson's system
(the log of the observations is given in Table 3). The seeing quality
ranged from 2.1\arcsec\ ($R$ images) to 2.8\arcsec\ ($V$ and $I$
images).

\begin{figure}
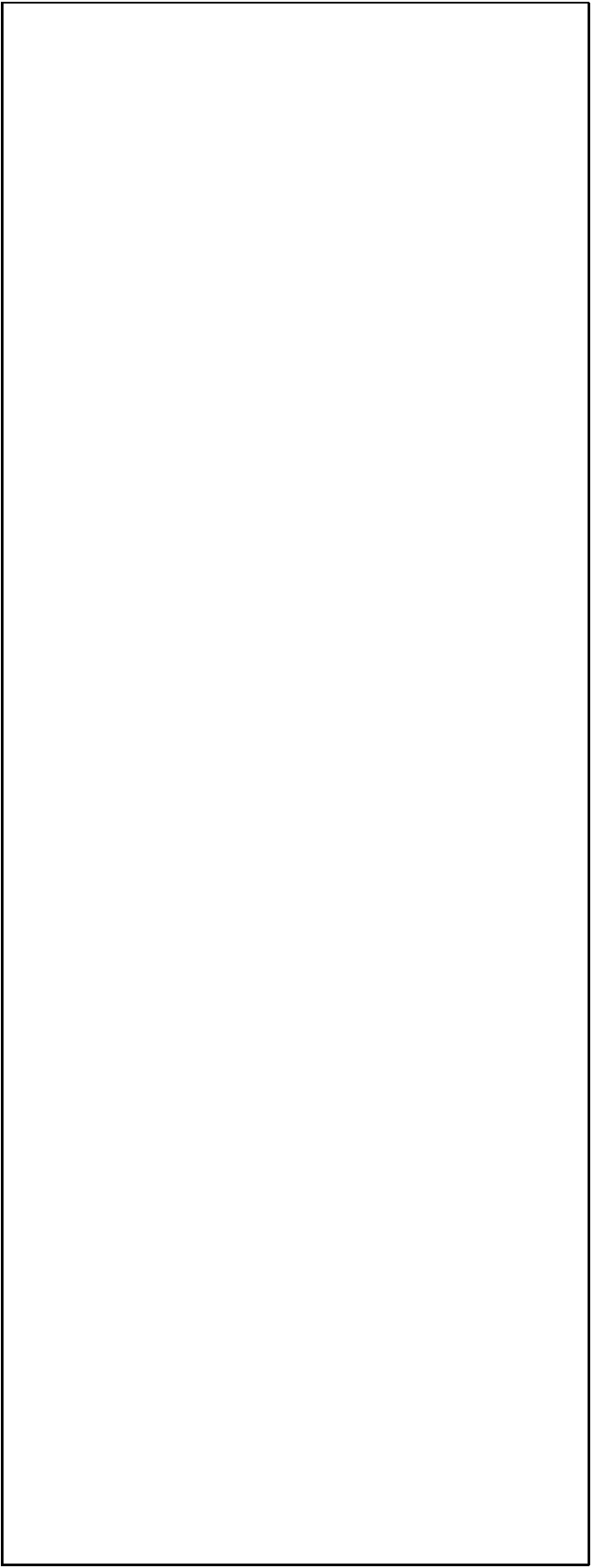

% Fig 2
\picplace{23.4 cm}
\caption[ ]{The $I$ and $B$ images of NGC 6181. The frame
is $108\arcsec\times108\arcsec$, north is up and east is
to the left.}
\end{figure}

Gray-scaled sky-subtracted $I$ and $B$ images are presented in
Fig. 2a and 2b; flux-calibrated isophotes in $B$ (21.8 $mag/arcsec^2$
and 25.3 $mag/arcsec^2$ for the innermost and the outermost isophotes
respectively with the step of 0.5 $mag/arcsec^2$) are shown in
Fig.2c. The $B$ and $V$ frames were calibrated by using 14 aperture
photoelectric measurements of NGC 6181 from Burstein et al. (1987).
The range of aperture radii is from 16\arcsec\ to 48\arcsec.
Zero-point magnitudes are obtained with accuracy better than 0.01 mag,
color terms are found to be negligible. The sky brightnesses are
estimated as 21.26 $mag/arcsec^2$ in $V$ and 22.12 $mag/arcsec^2$ in
$B$; these values coincide with mean sky brightnesses measured fifteen
years ago in the
Special Astrophysical Observatory at $z=60^o$ by Neizvestny (1981).
To check our $BV$ calibration, we compared multi-aperture
photoelectric data for NGC 6181 taken from the catalogue of Longo \&
Vaucouleurs (1983, 1985) (18 entities excluding old data of PET-54 and
BIG-51) with values simulated for the same apertures from our CCD
frames. In Fig. 3 one can see a rather good consistency with the
photoelectric data, even for large apertures. The calibration of $R$
and $I$ frames was indirect and less precise because standard stars
were not observed. Instead  $B-V$ colors for five faint stars in the
field of the galaxy were measured, and adopting them to be dwarfs we
ascribe them mean $V-R$ and $V-I$ colors in accordance with their
spectral types (Straizys 1977). The formal accuracy of the
calibration constants determined in such a way is 0.12 mag, but we
admit a possible systematic shift of our $R$ and $I$ magnitudes by up
to 0.3 mag.

\begin{figure}
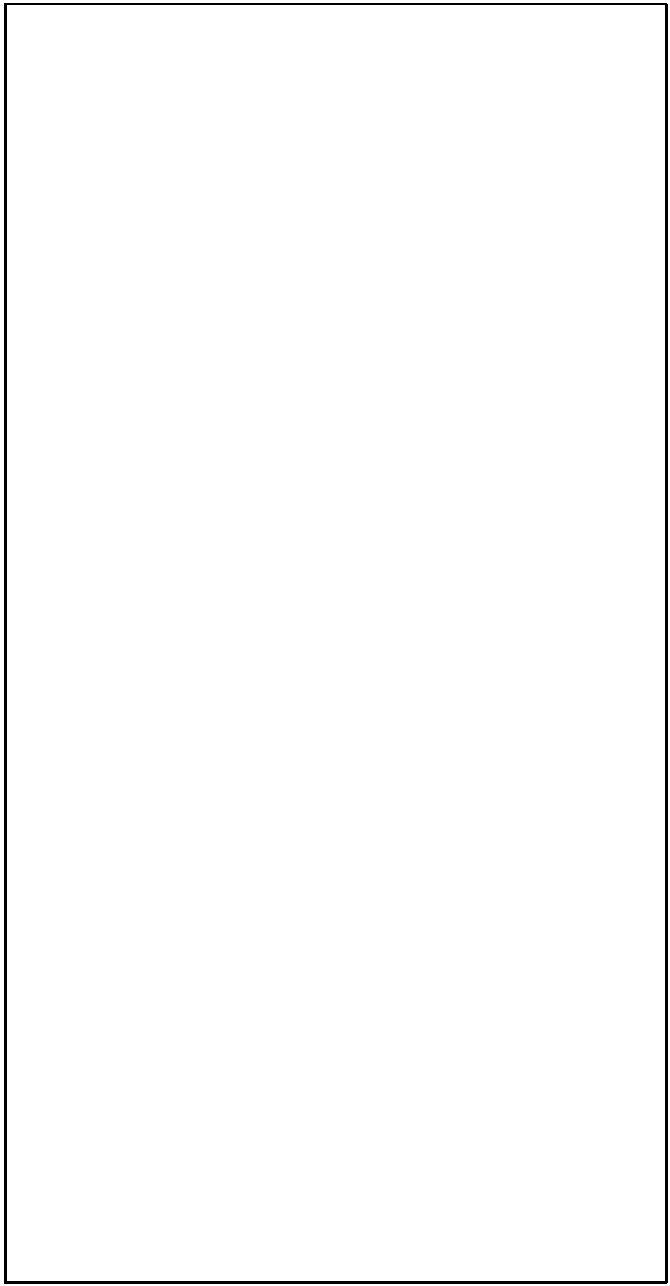

% Fig 3
\picplace{17.0 cm}
\caption[ ]{Comparison of published $V$ and $B$ multi-aperture photoelectric
data on NGC 6181 and present CCD observations.}
\end{figure}

\section{Photometric properties of NGC 6181}

A monochromatic image was constructed from the Perot-Fabry
observational data in order to see the morphology of the inner part of
NGC 6181. The H$_\alpha$ map of the central part of NGC 6181 (Fig. 1)
reveals two bright central sources none of which coincides with the
center of the isophotes in the continuum. Two faint tails of
H$_\alpha$ emission embrace the central continuum source which is
located in the area of very weak emission. Bearing in mind the absence
of radio emission from the NGC 6181 nucleus, one may conclude that the
nucleus of this galaxy is very quiescent.

The direct images obtained with the 1m telescope were used first of
all to find the precise position of the center of the galaxy in the
continuum. It was determined with respect to five nearby stars.
Location of the center was compared with the H$_\alpha$ distribution
and with the dynamical center position (see the next section). In
addition we tried to derive some surface brightness distribution
characteristics. Fig.2 demonstrates rather smooth image in $I$,
with a weak bar-like disk elongation in the inner part, and a more
clumpy image in $B$; prominent spiral arms extend up to the outermost
radii in all passbands, confirming the grand-design classification of
arms made by Elmegreen \& Elmegreen (1984).

\begin{figure}
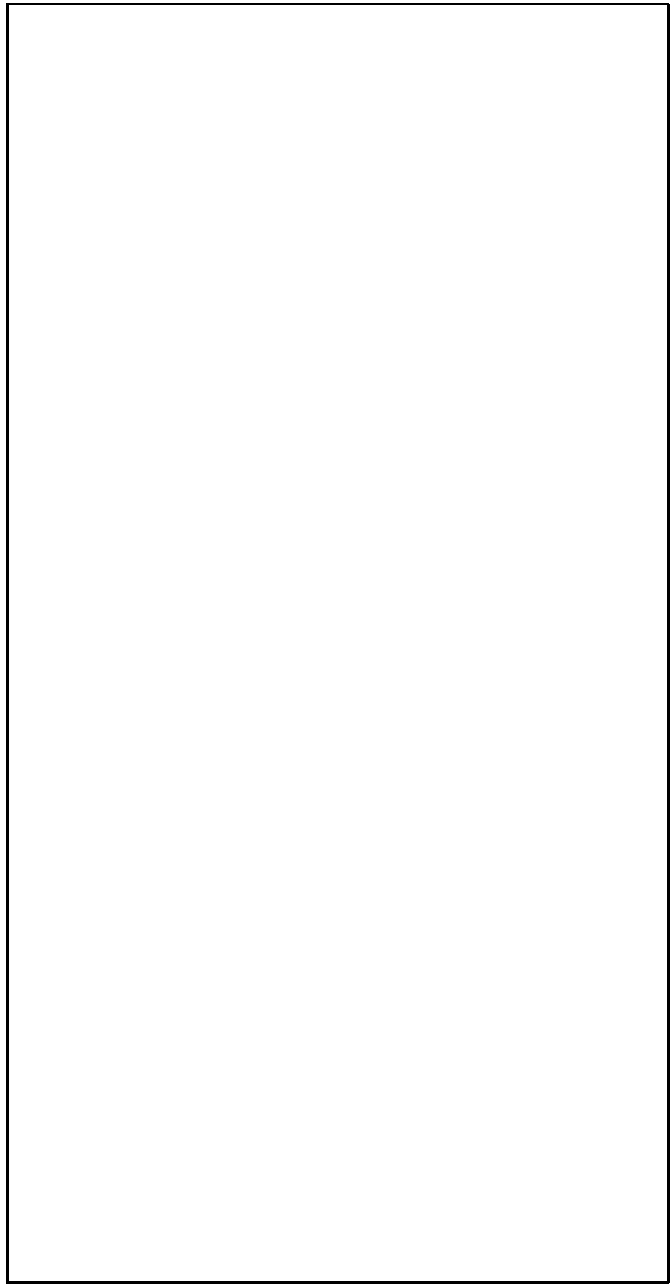

% Fig 4
\picplace{17.0 cm}
\caption[ ]{Radial distributions of azimuthally averaged surface brightness
in the $B$ and $V$ passbands. The straight lines represent the disk exponential
laws fitted to the radius range of 27\arcsec\ to 46\arcsec.}
\end{figure}

Fig. 4 presents $B$ and $V$ azimuthally averaged radial
surface brightness profiles
assuming $\it{PA}$ (line of nodes)=$173^o$ and inclination
$56^o$ in accordance with the velocity field analysis (see below). It
shows that this galaxy possesses a very compact bulge which  does not
affect light distributions beyond the radius  7\arcsec. In the range
12\arcsec--25\arcsec\ a brightness excess is noticeble over the simple
exponential law extrapolated from the outer parts; this excess is
reproduced in all four filters being the largest (0.15 mag)
in the $B$ passband. It seems that the radius of 25\arcsec\ is a
boundary between two disk subsystems. The disk scale measured in the
range 25\arcsec--45\arcsec\ for all four passbands slightly
decreases from blue to red (Fig. 5) being in general accordance with
earlier results of Elmegreen \& Elmegreen (1984) and Roth (1994).

\begin{figure}
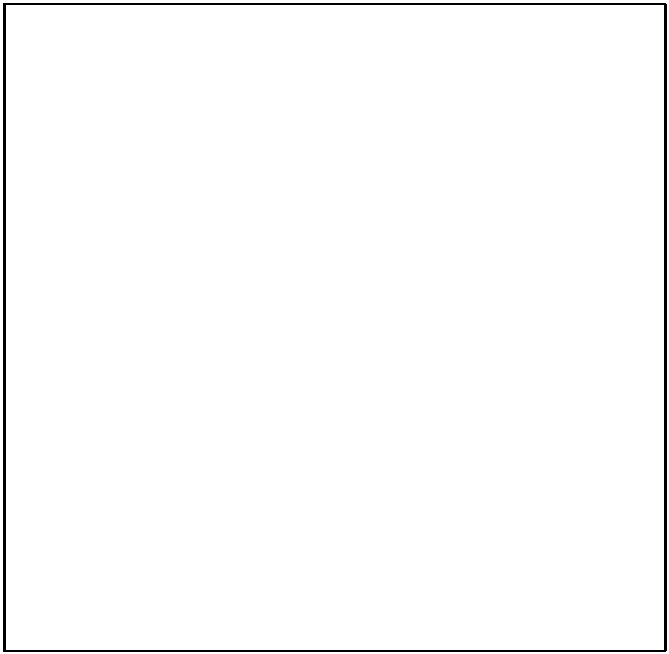

% Fig 5
\picplace{8.6 cm}
\caption[ ]{Disk scale variation with spectral range. Previous
published data are also plotted as comparison.}
\end{figure}

\begin{figure}
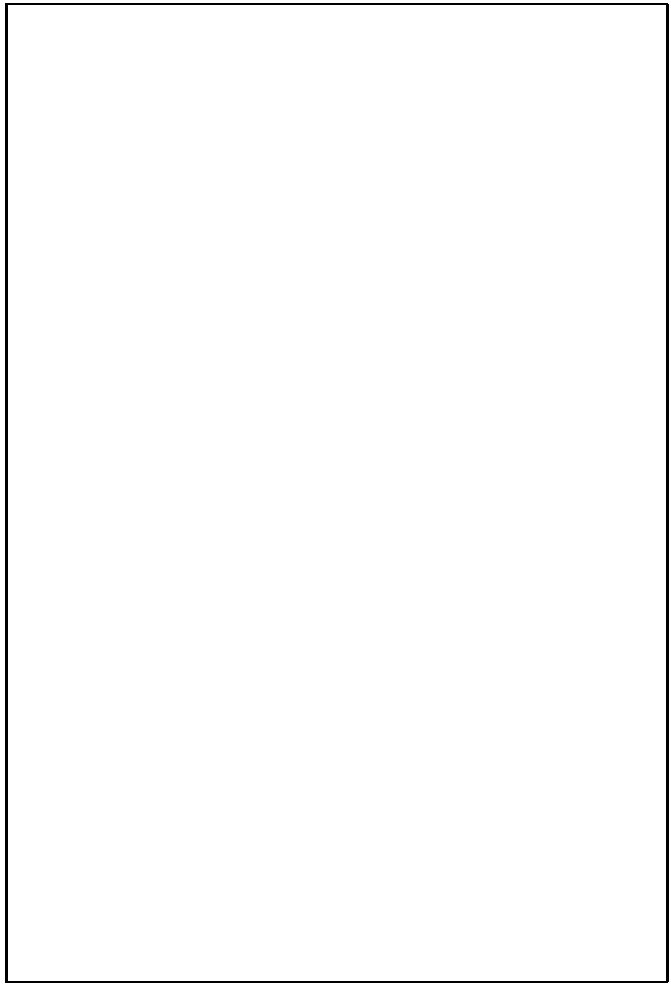

% Fig 6
\picplace{13.0 cm}
\caption[ ]{Color radial profiles averaged over azimuth in the galaxy plane.
The r.m.s. error of a single point is less than 0.05 mag}
\end{figure}

The azimuthally averaged color tends to be bluer up to $r\approx
36\arcsec$ and then some reddening occurs, but the slopes of these
trends are quite different for different colors (Fig. 6 and 7). In
Fig. 7 we try to compare radial color variations in NGC 6181 with
models for old stellar populations showing pure metallicity trend
(Worthey 1994) and with a mean observational sequence of galactic
colors (Buta \& Williams 1995), which is known to be defined mainly
by different present-time star formation rate. Color excesses expected
due to interstellar reddening in the Galaxy are also shown.
The reddening in $B-V$ at $r\approx 5\arcsec$ has an azimuthally
non-homogeneous character: it is a distinct, very red spot to the west
from the nucleus obviously related to a local dust concentration;
the optical characteristics of the dust may be unusual because
the spot is absent in $V-R$ and $V-I$ colors. The other color
variations seem to be rather azimuthally homogeneous. Comparison
of the observed and the expected color trends shows that between the
radii 7\arcsec\ and 25\arcsec, i. e. in the inner disk distinguished
by some brightness excess, the observed color variations may be
satisfactorily explained by variations of star formation intensity,
because the point grid is roughly parallel to the observational
sequence of integrated galactic colors. However the observed color
variations in the outer disk are more complicated and rather unusual --
especially for $r > 36\arcsec$, where the reddening of $B-V$ and $V-R$ occurs
under the constant $V-I$. This looks quite inexplicable in the frames of
simple effects which influence the color.

\begin{figure}
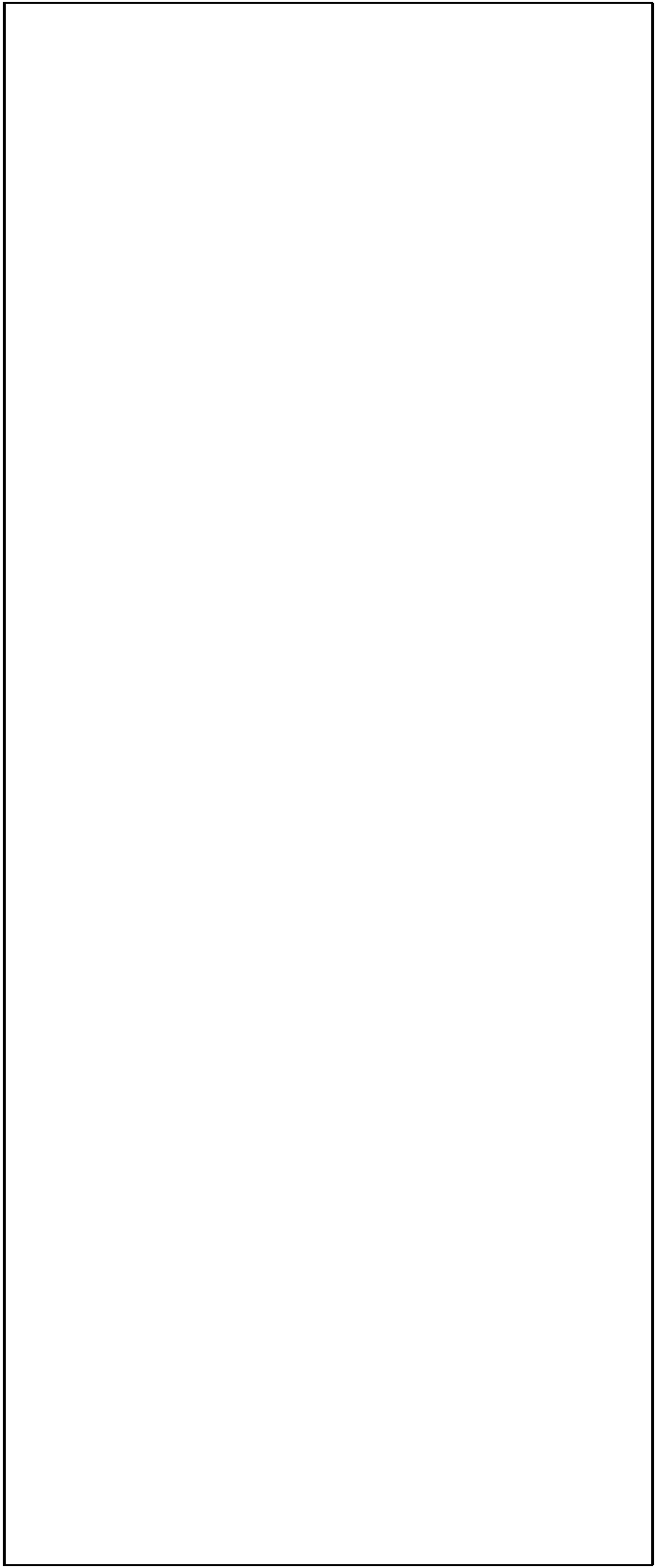

% Fig 7
\picplace{21.2 cm}
\caption[ ]{($V-R$, $B-V$) (a) and ($V-I$, $B-V$) (b) diagrams
for the radial color
variations in NGC 6181. Points are plotted through one arcsecond step.
Estimates of $V-R$ and $V-I$ have systematic shift (see the text).}

\end{figure}

\begin{figure}
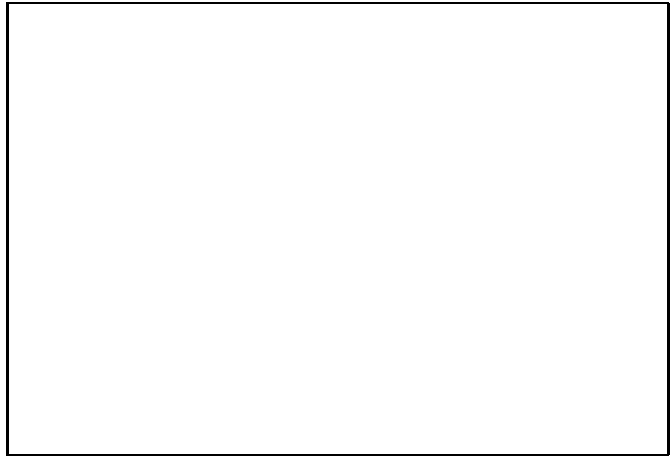

% Fig 8
\picplace{6.0 cm}
\caption[ ]{Variations of the orientation of the photometric major axis
along the radius. For $r<4\arcsec$ measurements taken for all passbands
were averaged, beyond this radius only $I$ isophotes were used.
The long-dashed line indicates the orientation of the outermost isophotes
according to our measurement of the SKYVIEW picture, the short-dashed line
shows the kinematical line of nodes.}
\end{figure}

The isophote form analysis (pure-ellipse fitting) was
carried out to check a possible deviation from axial symmetry. The
ellipticity between 1\arcsec\ and 7\arcsec\ from the center gradually
increases from 0.10 (bulge) to 0.35 -- a behavior which is quite
normal for a galaxy whose inclination is about $60^o$. The radial
dependence of $\it{PA}_0$ is presented in Fig. 8. We see an
unambiguous turn of isophotes in the very center of NGC 6181.
Measurements in all passbands show that at the radius of
2\arcsec--3\arcsec\ the position angle of the major axis is $+3^o$
with an uncertainty less than $1^o$, which differs by $\approx 5^o$
from the orientation of the outermost isophotes ($175^o$, Nilson 1973;
$178.4^o\pm0.5^o$, our measurement of the SKYVIEW isophote at the
$r=65\arcsec$). In the radius range 10\arcsec--20\arcsec\ isophotes are
also twisted by $10^o-13^o$, but in the opposite sense with respect to
the innermost region. Only beyond $r\approx 25\arcsec$, where the
radial brightness distribution follows a pure exponential law, the
isophote major axis becomes aligned with the line of nodes.

\section{Velocity field of the ionized gas in NGC 6181}

Fig. 9 presents the observed velocity field of NGC 6181.
It looks quite regular, with prominent signs of rotation. However in
the center of the galaxy a twist of the zero-velocity line is seen which
gives evidence for non-circular gas motions in this area in a
good agreement with the photometrical data (see below).

\begin{figure}
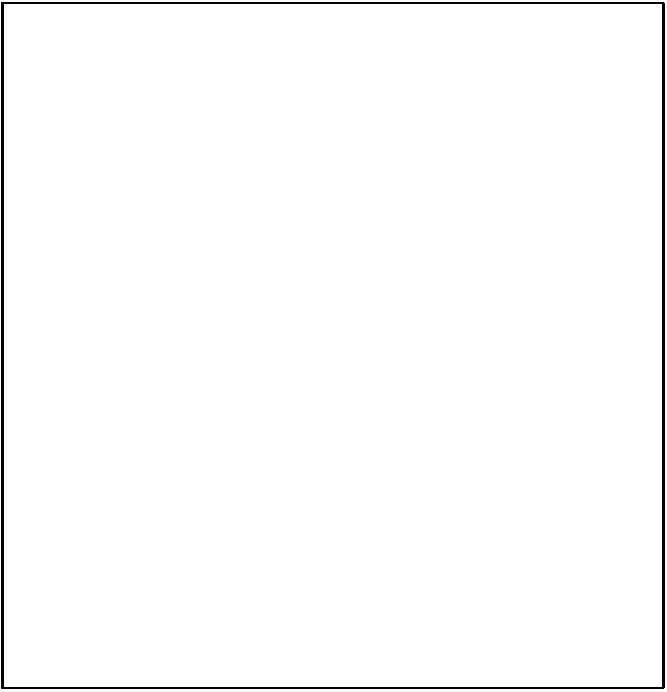

% Fig 9
\picplace{9.1 cm}
\caption[ ]{Isovelocity map of NGC 6181. Velocities are in $km\cdot s^{-1}$.
Cross marks the nucleus location.}
\end{figure}

A fit of the circular rotation model was made for the full velocity map,
with dimensions of $210\times185$ pixels
($147\arcsec\times130\arcsec$), or within a radius of about 75\arcsec\
(12.2 kpc) from the center. Special codes were written for data
processing. The galactic disk was supposed to be thin and flat, that is
it does not have any tilt of warp within the optical radius; so, we
looked for inclination $i$ and position angle of the line of nodes
$\it{MA}$ for the full radius range.  As a first step, we determined the
position of the dynamical center and a systemic velocity suggesting a
central symmetry of the velocity field. The dynamical center appears
to coincide with the center of broadband isophotes with an accuracy of
one pixel. The systemic velocity is found to be 2375 $km\cdot s^{-1}$ which
agrees with earlier determinations (Table 1). The dispersion of
systemic velocity values determined over all pairs of symmetrically
taken points of the galaxy is 16 $km\cdot s^{-1}$ which is close to our
accuracy of individual velocity determinations. Then we verified if
the whole line-of-sight velocity field can be fitted by a pure
circular rotation. The mean line-of-sight velocity residuals (r.m.s.)
were calculated for $30^o$ ranges of $i$ and $\it{MA}$; the minimum of
the velocity residual calculated over the total velocity field reveals
the true values of these parameters. This approach assumes that any
velocity field distortions, if they exist, are of local nature.

The agreement is found the best for the following orientation
parameters: $i_0=56^o$, $\it{MA}=173^o$, the $\it{MA}$ being more
strictly limited and $i_0$ having less accuracy.
If one compares these values with the photometric parameters of NGC
6181 -- for example, $i=61^o$ (Bottinelli et al. 1984) and
$\it{MA}=175^o$ (Nilson 1973) or with the data from the Table 1 -- it
becomes clear that the bulk of the gas in the galaxy rotates
circularly.

\begin{figure}
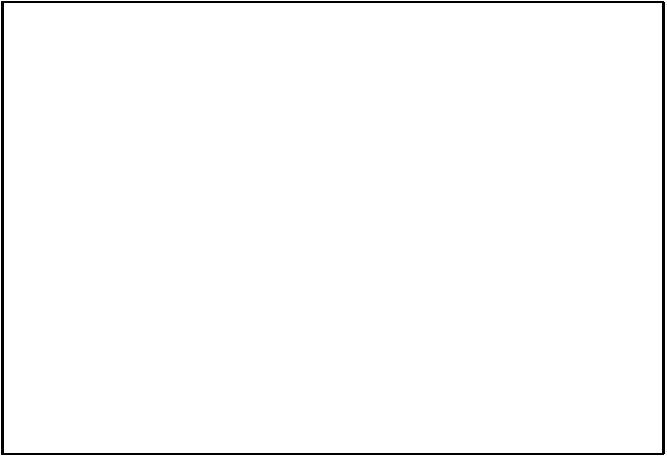

% Fig 10
\picplace{6.0 cm}
\caption[ ]{Azimuthally averaged rotation curve of NGC 6181 obtained under
the assumption of pure circular rotation. The inner region (${< 5\arcsec}$)
is excluded because of the presence of non circular motions in the center.
}
\end{figure}

Fig. 10 presents the azimuthally averaged rotation curve. For
$r>25\arcsec$ it is nicely flat at the level of 210 $km\cdot s^{-1}$ with
an r.m.s. error of individual points not worse than 5-7 $km\cdot s^{-1}$.
The maximal rotation velocity estimated from the width of the HI line
at 21 cm, $W50$, is 222 $km\cdot s^{-1}$ (LEDA Consultation), so our
rotation curve for NGC 6181 is in accordance with previously known
data.

Despite the generally good accordance between the observed velocity
field and the circular rotation model, there are three ranges of
radial distances where essential systemic deviations from a pure
circular rotation are detected with the regions of maximal deviations
at $r = 30\arcsec - 40\arcsec$ , $r \approx 12 \arcsec$ and $r< 3
\arcsec$. The first area is located near to the dynamical major axis
and coincides neither with spiral arms nor with bright HII
regions. However, the color profiles discussed in the previous section
demonstrate a turnover of color radial trends at this radius. In the
southern half of the galaxy this region is distinguished by the excess
of azimuthal velocity of order of 30--50 $km\cdot s^{-1}$, and in the
northern half of the galaxy there is a similar velocity depression of
about the same value. As these two areas are located symmetrically
with respect to the galactic center, this anomaly may be considered
rather as a kind of regular wave distortion of the velocity field than
as a local velocity anomaly.

\begin{figure}
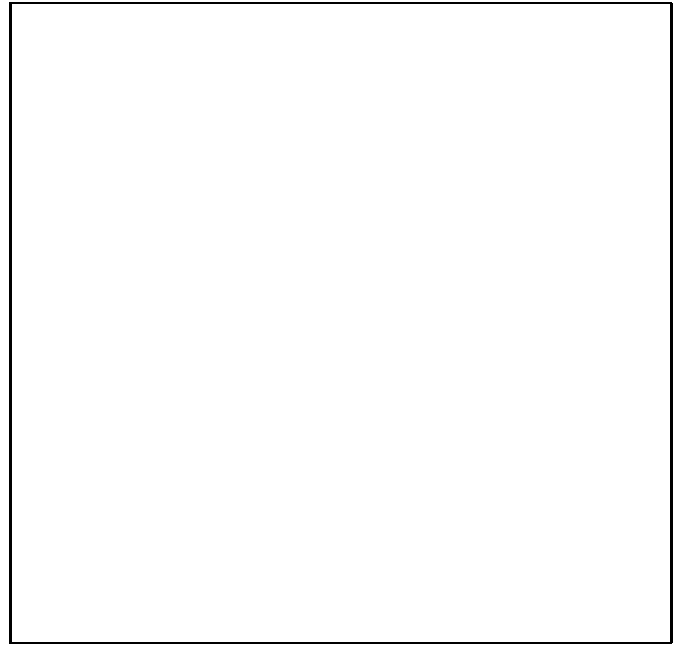

% Fig 11
\picplace{8.5 cm}
\caption[ ]{Residual velocities from pure circular rotation for the
inner part of the galaxy. Blank areas correspond to residual
velocities between --20 and +25 $km\cdot s^{-1}$; medium blackness to --40
to --20 $km\cdot s^{-1}$; darkest regions represent positive residuals up
to +70 $km\cdot s^{-1}$. The inner pair of shaded areas is related to the
central mini-bar. The most  remarkable extended shaded areas present
the ring-like zone of radial gas motions at  radius of about
10--15\arcsec. }
\end{figure}

The model velocity field calculated in the frame of pure circular rotation
with the parameters mentioned above has been subtracted from the
observed velocity field. The central part of the residual velocity
field is presented in Fig. 11. Here we see two halves of the ring-like
region where deviations from circular rotation model locally exceed
50 $km\cdot s^{-1}$. Being deprojected onto the plane of the galaxy, this
region looks like three quarters of a perfect circular ring with a
mean radius of 11\arcsec\ (about 1.8 kpc); the eastern half of the
ring has positive residual velocities up to 55 $km\cdot s^{-1}$, the
western part has negative ones, from --30 to --40 $km\cdot s^{-1}$. The
width of the ring is at least 5 pixels, which corresponds to 0.8 kpc.
The fact that the switching of residual velocity sign takes place near
the line of nodes implies that the residual velocities here are
mostly radial ones (here and below, we admit that the gas motion
is in the plane of the disk). Together with a circular deprojected
shape of the ring, it also gives evidences that the line of nodes of
this structure is close to the $\it{MA}$ of the global galactic disk
and that the ring lies exactly in the galactic plane being purely an
internal feature of the galactic gaseous subsystem. Stemming from
the slightly asymmetric minor-axis surface brightness profile of the bulge
one may conclude that the western half of the galactic disk is the nearest
one to us. In this case we may conclude that NGC 6181 possesses a trailing
spiral pattern and radially expanding gas motions in the ring.

It is worth noting that the ring of radially moving gas lies closer to
the center than the beginning of the well-defined spiral arms and does
not reveal itself in a brightness distribution. Nevertheless ionized
gas in the ring shows systemic velocity residuals of much higher
amplitude than in the bright spiral arms.

\begin{figure}
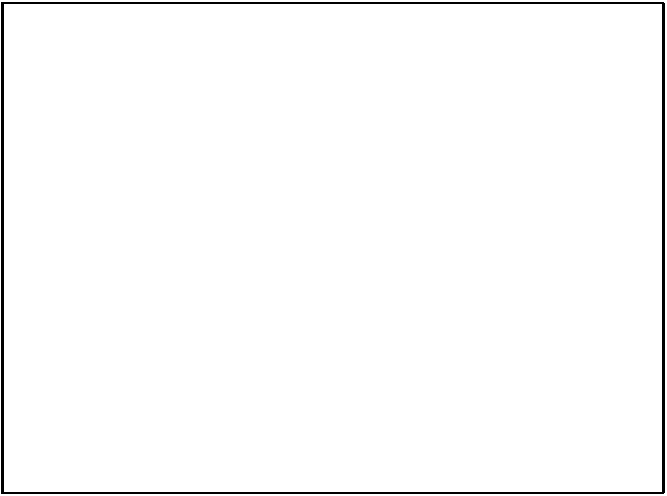

% Fig 12
\picplace{6.5 cm}
\caption[ ]{Azimuthal dependence of the line-of-sight velocity central
gradients within the radius range 1.8\arcsec -- 2.2\arcsec.
The solid curve represents a cosine law fitted by a least-square algorithm.
}
\end{figure}

In the very center of NGC 6181, within the region where the major axis
of continuum isophotes is twisted ($r<5\arcsec$), a clear sign of
elliptical gas rotation is seen in the two-dimensional velocity
field. Analysing an azimuthal dependence of central velocity
gradient, we find that the maximum of the cosine curve
computed by the least square approximation \\

\noindent
dv$_r$/dr = [44.2 cos($\it PA$ -- 325.5$^o$) -- 4.7]
$km\cdot s^{-1}\cdot arcsec^{-1}$ \\

\noindent
is shifted by about
$30^o$ relative to the line of nodes $\it{MA}=173^o$ (Fig. 12). Hence,
circumnuclear gas rotation in NGC 6181 parallel with the isophote
major axis twist gives strong evidence for the presence of a small bar
in the very center of the galaxy. So NGC 6181 may be applied to a
small number of known galaxies where nuclear bar reveals itself both
from photometric and kinematic data.

A physical connection between the central mini-bar and the ring-like
zone of gas expansion may be suspected. It follows from radial
velocities of gas in the ring deprojected onto the plane of the
galaxy, assuming that these motions are purely radial. Parameters of
the galactic plane orientation used for deprojection were taken from
the best fit model of circular rotation: $i_0=56^o$, $\it{MA}=173^o$.
It appeared that the radial velocity of the ring expansion varies
along the eastern half of the ring from 50 to 120 $km\cdot s^{-1}$ (Fig.
13), and the position angle of the maximum expansion velocity roughly
coincides with the position angle of the minimum of cosine curve
describing the azimuthal dependence of the central velocity gradient.
It gives some evidence that the position angle of the largest radial
velocities is related to the orientation of nuclear bar.

\begin{figure}
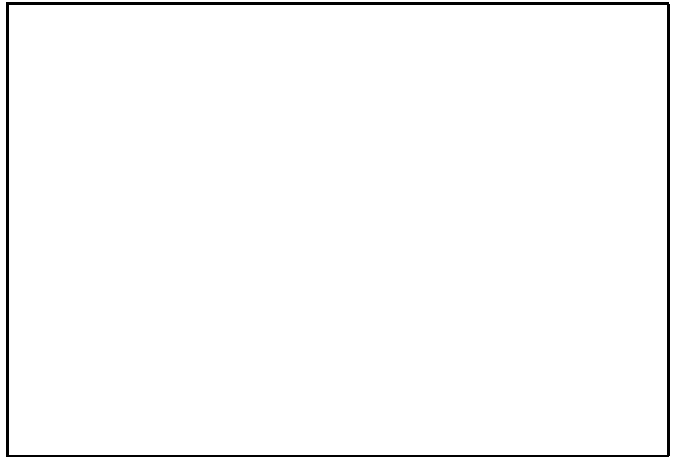

% Fig 13
\picplace{6.0 cm}
\caption[ ]{Azimuthal dependence of the gas radial velocities in the
ring-like zone with the radius of 12\arcsec: residual velocities presented
in Fig. 11 are deprojected onto the galactic plane under the assumption
of pure radial motions with the galactic plane orientation parameters
$\it i$ = 56$^o$ and $\it {MA}$=173$^o$.
}
\end{figure}

Kinematically distinct, ring-like inner regions of systemic
radial motions of gas, similar to what we observe in NGC 6181, were not known
yet. The only analogy, which can be mentioned, is the famous "3 kpc
arm" in our Galaxy: it possesses radial velocities of about 100--150
$km\cdot s^{-1}$ and is probably connected with a triaxial structure of the
Galactic center. The other radially expanding gaseous ring, found
in NGC 4725 (Buta 1988), has a radius of 10--13 kpc being a structure
of a quite different scale.

Note, that morphologically distinguished nuclear rings, which reveal themselves
as zones of brightness, not of velocity, excess, often
accompany nuclear bars (Buta \& Crocker 1993). Numerical simulations
confirm that gaseous subsystems may give ring-like response to a
triaxial potential form (Combes \& Gerin 1985). So what we found may
be considered as a kinematical counterpart of such structures as nuclear
rings, related to general disk structure. In this case, as in
the case of our Galaxy, it is not necessary to interpret radial gas
velocities in the ring-like zone as an evidence of its real expansion due to
some explosion event: there is no hint of the presence of a shock front
or enhanced star formation in front of the ring or in the ring itself.
A more realistic explanation is that we observe here an unusually large
amplitude of hydrodynamical oscillations of gas velocities associated
with the density waves which penetrate deep into the inner part of the
disk (Fridman et al., in preparation).

The other feature of the velocity field of NGC 6181 is a
multicomponent structure of emission line profiles in some
HII regions of the disk. We performed Gauss analysis of
two-component emission line profiles for the central part of the
galaxy $64\times64$ pixels, or $45\arcsec\times45\arcsec$. The primary
-- more strong and everywhere narrow -- component reveals a velocity
field which excellently agrees with the field obtained in the previous
analysis: a general circular rotation, elliptical gas motions in the
center and ring-like zone of radial gas motions. The secondary, more
weak and broad component (with gas velocity dispersion up to
200 $km\cdot s^{-1}$) appears only in the four brightest HII regions; it is
absolutely absent in the ring-like zone of the radial gas motions. The
difference between the "first" and the "second" velocity component
averaged over the total region of the galaxy is zero, but for the
bright HII region, nearest to the dynamical center, there exists a switch
of velocity difference sign between the northern and the southern
halves. It allows to suspect that we deal with a proper rotation
of a giant star formation site.

\section{Concluding remarks}

Measurements of the line-of-sight velocity field for gas emission in
the giant Sc galaxy NGC 6181 have allowed us to reveal some rare
phenomena. The most unusual of them is the presence of a ring-like
region with striking systemic deviations from the general circular
rotation which may be interpreted as a strong gas radial motion. More
elaborate interpretation which connects this motion with the presence
of hydrodynamical 3D oscillations of gas velocities related to density
wave phenomena will be given elsewhere (Fridman et al., in
preparation). A small nuclear bar is also found in the galaxy both by
kinematic and photometric methods.

\begin{acknowledgements}
We are very grateful to the observers of the Special Astrophysical Observatory
RAS -- S.N. Dodonov, S.V. Drabek, and V.V. Vlasiuk assisting us at the
6 m telescope, and V.O. Chavushyan and S.K. Balayan which obtained
the data at the 1 m telescope. During the data analysis we have
used the Lyon-Meudon Extragalactic Database (LEDA) supplied by the
LEDA team at the CRAL-Observatoire de Lyon (France) and of the NASA/IPAC
Extragalactic Database (NED) which is operated by the Jet Propulsion
Laboratory, California Institute of Technology, under contract with
the National Aeronautics and Space Administration. We acknowledge also
an anonymous referee for the very useful comments and advices.
\end{acknowledgements}

\end{document}